\documentclass[aps,prab,reprint,groupedaddress]{revtex4-2}

\usepackage[english]{babel}
\usepackage{siunitx}
\usepackage{lipsum}
\usepackage{tikz}
\usetikzlibrary{calc, shapes, decorations.pathmorphing, positioning, arrows, arrows.meta, external}
\usepackage{pgfplots}
\usepgfplotslibrary{groupplots}
\pgfplotsset{
	compat=1.13,
}
\usepackage{amsmath}
\usepackage[siunitx]{circuitikz}
\usepackage{csquotes}

\graphicspath{{tikz/}}

\definecolor{TolRainbow_01}{HTML}{E8ECFB}
\definecolor{TolRainbow_02}{HTML}{D9CCE3}
\definecolor{TolRainbow_03}{HTML}{D1BBD7}
\definecolor{TolRainbow_04}{HTML}{CAACCB}
\definecolor{TolRainbow_05}{HTML}{BA8DB4}
\definecolor{TolRainbow_06}{HTML}{AE76A3}
\definecolor{TolRainbow_07}{HTML}{AA6F9E}
\definecolor{TolRainbow_08}{HTML}{994F88}
\definecolor{TolRainbow_09}{HTML}{882E72}
\definecolor{TolRainbow_10}{HTML}{1965B0}
\definecolor{TolRainbow_11}{HTML}{437DBF}
\definecolor{TolRainbow_12}{HTML}{5289C7}
\definecolor{TolRainbow_13}{HTML}{6195CF}
\definecolor{TolRainbow_14}{HTML}{7BAFDE}
\definecolor{TolRainbow_15}{HTML}{4EB265}
\definecolor{TolRainbow_16}{HTML}{90C987}
\definecolor{TolRainbow_17}{HTML}{CAE0AB}
\definecolor{TolRainbow_18}{HTML}{F7F056}
\definecolor{TolRainbow_19}{HTML}{F7CB45}
\definecolor{TolRainbow_20}{HTML}{F6C141}
\definecolor{TolRainbow_21}{HTML}{F4A736}
\definecolor{TolRainbow_22}{HTML}{F1932D}
\definecolor{TolRainbow_23}{HTML}{EE8026}
\definecolor{TolRainbow_24}{HTML}{E8601C}
\definecolor{TolRainbow_25}{HTML}{E65518}
\definecolor{TolRainbow_26}{HTML}{DC050C}
\definecolor{TolRainbow_27}{HTML}{A5170E}
\definecolor{TolRainbow_28}{HTML}{72190E}

\pgfplotscreateplotcyclelist{TolRainbow7 cycle}{%
  {TolRainbow_09},
  {TolRainbow_10},
  {TolRainbow_14},
  {TolRainbow_15},
  {TolRainbow_17},
  {TolRainbow_18},
  {TolRainbow_26},
}

\pgfplotscreateplotcyclelist{TolRainbow20 cycle}{%
  {TolRainbow_02},
  {TolRainbow_04},
  {TolRainbow_05},
  {TolRainbow_07},
  {TolRainbow_08},
  {TolRainbow_09},
  {TolRainbow_10},
  {TolRainbow_11},
  {TolRainbow_13},
  {TolRainbow_14},
  {TolRainbow_15},
  {TolRainbow_16},
  {TolRainbow_17},
  {TolRainbow_18},
  {TolRainbow_20},
  {TolRainbow_22},
  {TolRainbow_24},
  {TolRainbow_26},
  {TolRainbow_27},
  {TolRainbow_28},
}

\definecolor{tolmint}{HTML}{44bb99}
\definecolor{tolpear}{HTML}{bbcc33}
\definecolor{tololive}{HTML}{aaaa00}
\definecolor{tolyellow}{HTML}{eedd88}
\definecolor{tolorange}{HTML}{ee8866}
\definecolor{tolpink}{HTML}{FFAABB}
\definecolor{tolcyan}{HTML}{99ddff}

\begin{document}

\title{Demonstration of electron cooling using a pulsed beam from an electrostatic electron cooler}

% repeat the \author .. \affiliation  etc. as needed
% \email, \thanks, \homepage, \altaffiliation all apply to the current
% author. Explanatory text should go in the []'s, actual e-mail
% address or url should go in the {}'s for \email and \homepage.
% Please use the appropriate macro foreach each type of information

% \affiliation command applies to all authors since the last
% \affiliation command. The \affiliation command should follow the
% other information
% \affiliation can be followed by \email, \homepage, \thanks as well.
\author{M.~W.~Bruker}\email[]{bruker@jlab.org}
\affiliation{Thomas Jefferson National Accelerator Facility, Newport News, Virginia 23606, USA}
\author{S.~Benson}
\affiliation{Thomas Jefferson National Accelerator Facility, Newport News, Virginia 23606, USA}
\author{A.~Hutton}
\affiliation{Thomas Jefferson National Accelerator Facility, Newport News, Virginia 23606, USA}
\author{K.~Jordan}
\affiliation{Thomas Jefferson National Accelerator Facility, Newport News, Virginia 23606, USA}
\author{F.~Ma}
\altaffiliation[Also at: ]{University of Chinese Academy of Sciences, Beijing 100049, China}
\affiliation{Institute of Modern Physics, Chinese Academy of Sciences, Lanzhou 730000, China}
%\altaffiliation{University of Chinese Academy of Sciences, Beijing 100049, China}
\author{J.~Li}
\affiliation{Institute of Modern Physics, Chinese Academy of Sciences, Lanzhou 730000, China}
\author{X.~M.~Ma}
\affiliation{Institute of Modern Physics, Chinese Academy of Sciences, Lanzhou 730000, China}
\author{L.~J.~Mao}\email[]{maolijun@impcas.ac.cn}
\affiliation{Institute of Modern Physics, Chinese Academy of Sciences, Lanzhou 730000, China}
\author{T.~Powers}
\affiliation{Thomas Jefferson National Accelerator Facility, Newport News, Virginia 23606, USA}
\author{R.~Rimmer}
\affiliation{Thomas Jefferson National Accelerator Facility, Newport News, Virginia 23606, USA}
\author{T.~Satogata}
\affiliation{Thomas Jefferson National Accelerator Facility, Newport News, Virginia 23606, USA}
\author{X.~P.~Sha}
\altaffiliation[Also at: ]{University of Chinese Academy of Sciences, Beijing 100049, China}
\affiliation{Institute of Modern Physics, Chinese Academy of Sciences, Lanzhou 730000, China}
%\altaffiliation{University of Chinese Academy of Sciences, Beijing 100049, China}
\author{A.~Sy}
\affiliation{Thomas Jefferson National Accelerator Facility, Newport News, Virginia 23606, USA}
\author{M.~T.~Tang}
\altaffiliation[Also at: ]{University of Chinese Academy of Sciences, Beijing 100049, China}
\affiliation{Institute of Modern Physics, Chinese Academy of Sciences, Lanzhou 730000, China}
\author{H.~Wang}\email[]{haipeng@jlab.org}
\affiliation{Thomas Jefferson National Accelerator Facility, Newport News, Virginia 23606, USA}
\author{S.~Wang}
\affiliation{Thomas Jefferson National Accelerator Facility, Newport News, Virginia 23606, USA}
\author{J.~C.~Yang}
\affiliation{Institute of Modern Physics, Chinese Academy of Sciences, Lanzhou 730000, China}
\author{X.~D.~Yang}
\affiliation{Institute of Modern Physics, Chinese Academy of Sciences, Lanzhou 730000, China}
\author{H.~Zhang}
\affiliation{Thomas Jefferson National Accelerator Facility, Newport News, Virginia 23606, USA}
\author{Y.~Zhang}
\affiliation{Thomas Jefferson National Accelerator Facility, Newport News, Virginia 23606, USA}
\author{H.~Zhao}
\altaffiliation[Currently at: ]{Brookhaven National Lab, Upton, New York 11973, USA}
\affiliation{Institute of Modern Physics, Chinese Academy of Sciences, Lanzhou 730000, China}
\author{H.~W.~Zhao}
\affiliation{Institute of Modern Physics, Chinese Academy of Sciences, Lanzhou 730000, China}

\date{\today}

% Notice to publisher
% Notice: Authored by Jefferson Science Associates, LLC under U.S. DOE Contract No. DE-AC05-06OR23177. The U.S. Government retains a non-exclusive, paid-up, irrevocable, world-wide license to publish or reproduce this manuscript for U.S. Government purposes.

\begin{abstract}

Cooling of hadron beams is critically important in the next generation of hadron storage rings for delivery of unprecedented performance. One such application is the electron-ion collider presently under development in the US.
The desire to develop electron coolers for operation at much higher energies than previously achieved necessitates the use of radio-frequency (RF) fields for acceleration as opposed to the conventional, electrostatic approach.
While electron cooling is a mature technology at low energy utilizing a DC beam, RF acceleration requires the cooling beam to be bunched, thus extending the parameter space to an unexplored territory.
It is important to experimentally demonstrate the feasibility of cooling with electron bunches and further investigate how the relative time structure of the two beams affects the cooling properties; thus, a set of four pulsed-beam cooling experiments was carried out by a collaboration of Jefferson Lab and Institute of Modern Physics (IMP).

The experiments have successfully demonstrated cooling with a beam of electron bunches in both the longitudinal and transverse directions for the first time. We have measured the effect of the electron bunch length and longitudinal ion focusing strength on the temporal evolution of the longitudinal and transverse ion beam profile and demonstrate that if the synchronization can be accurately maintained, the dynamics are not adversely affected by the change in time structure.
\end{abstract}

\maketitle

\section{\label{sec:introduction}Introduction}

Electron cooling has become one of the most effective methods for increasing the phase space density of stored ion beams through their interaction with an electron beam copropagating at the same average velocity~\cite{Budker1967}. The first electron cooling experiment was successfully carried out at NAP-M (Novosibirsk) with nonrelativistic protons in 1974~\cite{Budker1976}. After decades of development, the electron cooling method has found a wide range of applications in several low- and medium-energy proton and ion storage rings. It has since become desirable to extend the method into the high-energy range of \SI{50}{\mega\electronvolt}, which will enable a high luminosity at future facilities such as the EIC~\cite{eic}.
However, most existing electron coolers are based on DC electron beams accelerated by electrostatic high voltage.
The highest-voltage cooler so far was successfully operated at FNAL at an energy of \SI{4.3}{\mega\electronvolt}~\cite{nagaitsev}.
Due to the technical limitations of high-voltage acceleration, providing cooling beams at much higher energies necessitates RF acceleration and thus the use of bunched electron beams.
A collaboration between Jefferson Lab (USA) and Institute of Modern Physics (IMP, China) was established in 2012 to conduct precursory bunched cooling experiments aimed at demonstrating the feasibility of such a scheme and investigating its beam-dynamical implications.
According to previous simulation results, the bunched electron beam cooling dynamics are different from those typically obtained with DC beams \cite{ZHAO2018219, PhysRevAccelBeams.21.023501}.
Because a dedicated facility for accelerating bunched electron beams for cooling purposes did not exist at the time, the availability and flexibility of the DC cooling setup installed in the ion ring CSRm at IMP led to the decision to add a pulsing option to the existing facility.
The first experiment, which took place in 2016, was the first demonstration of cooling with a pulsed beam \cite{Mao:COOL2017-TUP15, Zhang:IPAC2018-TUPAL069}, albeit with bunch lengths of several meters.
Four experiments were performed in total (in 2016, 2017, 2018, and 2019), the latter aiming at improving the data quality and addressing unresolved questions. Unlike the pioneer experiment, the most recent data set includes cooling of all stored bunches to improve statistics.
The beam current is also boosted by applying DC cooling during accumulation.
The first electron cooler based on RF acceleration of short ($\propto \si{\centi\meter}$) electron bunches was proposed in 2013 \cite{fedotov2013} and recently commissioned at RHIC with an electron energy of \SI{2}{\mega\electronvolt}~\cite{PhysRevLett.124.084801}.

We describe the pulsed-electron-beam cooling facility we set up at IMP and present the results of the most recent cooling experiment with ${}^{86}\mathrm{Kr}^{25+}$ ions at an energy of \SI{5}{\mega\electronvolt/nucleon}. 

\section{\label{sec:setup}Experimental setup}

\subsection{Storage ring layout}

CSRm is a racetrack-shaped synchrotron with a circumference of \SI{161}{\meter}. An electron cooler with an effective length of \SI{3.4}{\meter} and an RF cavity are placed in the dispersion-free sections.
While the ring can be operated with a coasting beam, the cavity provides the option to create bunches of adjustable length.
The details of the facility are described in \cite{XIA200211}.

\subsection{Generation of the pulsed cooling beam}

\begin{figure}[htb]
\includegraphics{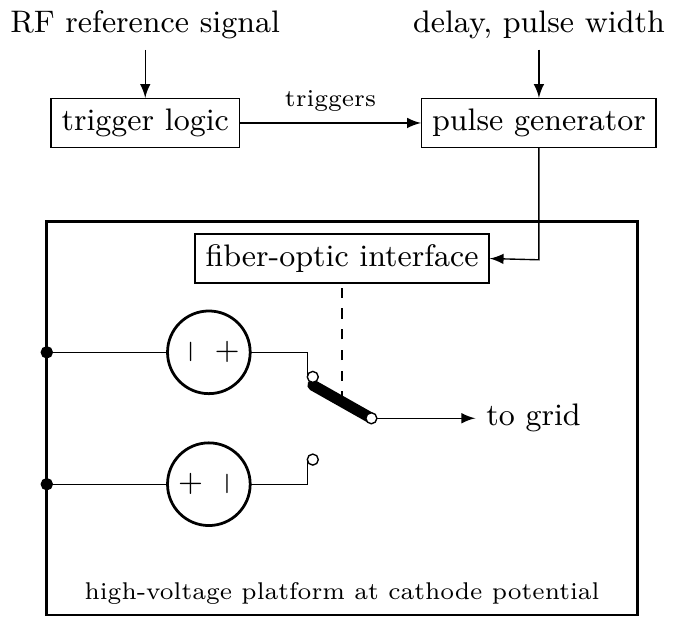}
% \tikzsetnextfilename{fig-pulsed-gun}
% \begin{tikzpicture}
% [
% 	transform shape,
% ]
% \draw (0, 0) node[cute spdt up, xscale=-1] (Sw) {};
% \draw (Sw.out 1) -- ++(0, 0.25) -- ++(-0.5, 0) to[american voltage source]  ++(-1, 0) coordinate (gnd1);
% \draw (gnd1) to[short, -*] (-3, 0 |- gnd1);
% \draw (Sw.out 2) -- ++(0, -0.25) -- ++(-0.5, 0) to[american voltage source, invert] ++(-1, 0) coordinate (gnd2);
% \draw (gnd2) to[short, -*] (-3, 0 |- gnd2);
% \draw[-latex] (Sw.in) -- ++(1.0, 0) node[anchor=west] {to grid};
% \draw [dashed] (Sw.mid) -- ++(0, 1.2) node[draw, solid, anchor=south] (fiberinter) {fiber-optic interface};
% 
% % high-voltage platform
% \draw [solid, thick] (-3, -2) rectangle (3, 2);
% \draw (0, -2) node[anchor=south, font=\scriptsize] {high-voltage platform at cathode potential};
% 
% \draw (-2, 4)
% 	node (rf) {RF reference signal};
% \draw (-2, 3)
% 	node [draw] (trigger) {trigger logic};
% \draw (2, 3)
% 	node [draw] (box) {pulse generator};
% \draw (2, 4)
% 	node (settings) {delay, pulse width};
% \draw[-latex] (trigger.east) -- (box.west)
% 	node [above, font=\scriptsize, midway] {triggers};
% \draw[-latex] (box.south) -- (box.south |- fiberinter.mid) -- (fiberinter.east);
% \draw[-latex] (settings.south) -- (box.north);
% \draw[-latex] (rf.south) -- (trigger.north);
% 
% \end{tikzpicture}
\caption{\label{fig:pulsed_gun}Schematic model of the electron pulse synchronization setup.}
\end{figure}

\begin{figure}[htb]
\includegraphics{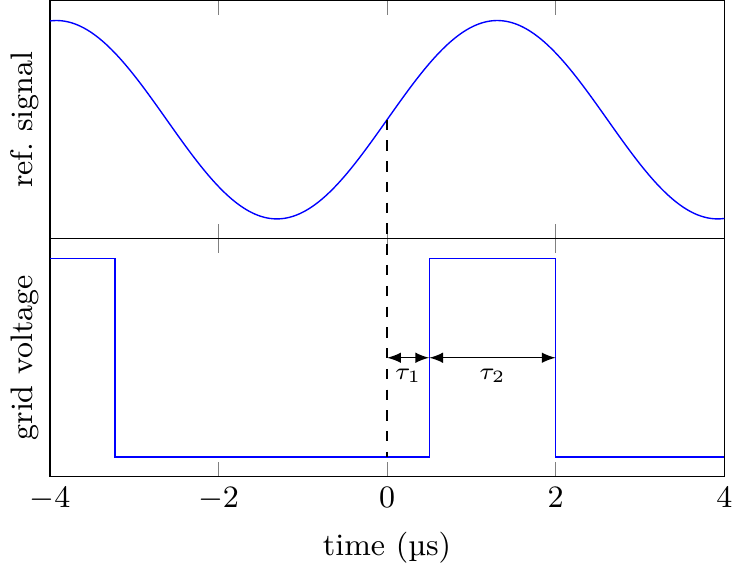}
% \tikzsetnextfilename{fig-delay-setup}
% \begin{tikzpicture}
% [
% ]
% \begin{groupplot}
% [
% 	group style={
% 		group size=1 by 2,
% 		x descriptions at=edge bottom,
% %		y descriptions at=edge left,
% 		vertical sep=0mm,
% 	},
% 	width=240pt,
% 	height=4cm,
% 	xlabel={time (\si{\micro\second})},
% 	xmin=-4, xmax=4,
% 	ytick=\empty,
% ]
% \nextgroupplot[ylabel={ref.~signal}]
% \addplot+[mark=none, domain=-4.0:4.0, samples=1000] {sin(deg(x*6.28*0.191372))};
% \draw (axis cs:0, 0) coordinate (origin);
% %\draw [solid] (axis cs:-4, 0) -- (axis cs:4, 0);
% \nextgroupplot[ylabel={grid voltage}]
% \addplot+[mark=none] coordinates {(-4, 1) (-3.2254, 1) (-3.2254, -1) (0.5, -1) (0.5, 1) (2, 1) (2, -1) (4, -1)};
% \draw (axis cs:0, -1) coordinate (foo);
% \draw (axis cs:0, 0) coordinate (bar);
% \draw (axis cs:0.5, 0) coordinate (baz);
% \draw (axis cs:2, 0) coordinate (qux);
% \end{groupplot}
% \draw [dashed] (origin) -- (foo);
% \draw [latex-latex] (bar) -- (baz) node[below, midway, font=\scriptsize] {$\tau_1$};
% \draw [latex-latex] (baz) -- (qux) node[below, midway, font=\scriptsize] {$\tau_2$};
% \end{tikzpicture}
\caption{\label{fig:delay_setup}Sketch of the fixed phase relation between the RF reference signal and the electron gun grid voltage. The delay $\tau_1$ and the bunch length $\tau_2$ are set independently.}
\end{figure}

The electron beam used for the cooling experiment is generated by the conventional magnetized electron cooler installed in the CSRm ring~\cite{BOCHAROV2004144}.
The existing electron source was modified to enable synchronization with the stored ion bunches in the following way:
The current emitted by the thermionic electron gun is a function of the voltage applied to the grid electrode. For normal operation, this voltage is set to a positive value that corresponds to a certain desired current, whereas it can be set to a negative value to shut off the electron beam completely.
By rapidly switching between these two voltages with a solid-state switch, we can generate rectangular electron pulses while leaving the other properties of the cooler essentially unmodified.
Using a reference signal generated by the RF system of the ring, these pulses are synchronized to the RF buckets, ensuring a stable temporal overlap between the electron bunches and the ion bunches.
The pulser setup is shown schematically in Fig.~\ref{fig:pulsed_gun} and the resulting voltage waveform in Fig.~\ref{fig:delay_setup}.

Because of the capacitance of the grid electrode and its cable, the currents flowing to charge and discharge the electrode with the desired high slew rate result in power dissipation in the switching element, which limits the available voltage and/or switching frequency.
Unlike a DC beam, which makes full use of energy recovery by default, a pulsed beam also loads the acceleration voltage supply of the cooler. The supply would therefore need to be bypassed at high frequencies to prevent excessive supply droop if a high peak current were to be extracted.
The beam parameters are a compromise to ensure that both the revolution frequency and the required voltages are well within the limits of the switching hardware \footnote{DEI PVX-4150}.
The choice of ${}^{86}\mathrm{Kr}^{25+}$ ions at an energy of $\SI{5}{\mega\electronvolt/nucleon}$, though far away from the properties of high-energy proton beams, is motivated by the use of this beam by existing users of the facility. Table~\ref{tab:all_params} lists the parameters used.

\begin{table}[!htbp]%[H] add [H] placement to break table across pages
\caption{Beam and instrumentation parameters\label{tab:all_params}}
\begin{ruledtabular}
\begin{tabular}{lc}
\textbf{ion beam} & \\
\hline \\ [-1.5ex]
particle type & ${}^{86}\mathrm{Kr}^{25+}$ \\
beam current & $< \SI{100}{\micro\ampere}$ \\
rest mass & \SI{930.5}{MeV/nucleon} \\
kinetic energy & \SI{5.0}{MeV/nucleon} \\
$\beta$ & 0.103 \\
$\gamma$ & 1.005 \\
revolution frequency & \SI{191.5}{\kilo\hertz} \\
harmonic number & 2 \\
RF voltage & 0.6--\SI{2}{\kilo\volt} \\ [1ex]
\hline
\textbf{electron cooler} & \\
\hline \\ [-1.5ex]
acceleration voltage & \SI{2.7}{\kilo\volt} \\
positive grid voltage & \SI{50}{\volt} \\
negative grid voltage & \SI{-551}{\volt} \\
peak current & \SI{30}{\milli\ampere}
\end{tabular}
\end{ruledtabular}
\end{table}

\subsection{Measurement of beam properties}

Depending on the available beam instrumentation devices, there are multiple independent ways to gain information about the cooling process.
The usual way to determine the longitudinal momentum distribution is by measuring the revolution peaks and their respective synchrotron sidebands from a Schottky pickup with a spectrum analyzer~\cite{forck_schottky}.
However, the cooling time in this experiment is short ($\approx \SI{1}{\second}$), and spectra with sufficient resolution and signal-to-noise ratio cannot be readily obtained from the available hardware. We therefore only use the spectrum from the Schottky pickup to determine the synchrotron frequency, which is a more dependable measurement of the RF voltage than the setting in the RF hardware itself.

Instead, the longitudinal cooling process is observed by measuring the temporal bunch profile using a beam position monitor (BPM) that is mounted outside the cooling region and thus only detects the ion bunches. The electron signal is measured by a similar BPM near the electron collector.
Since the distance between the two BPMs---and, thus, the time-of-flight difference---is known, the longitudinal overlap of the bunches can be measured and adjusted.

To remove the transverse information from the BPM signal, the signals from two opposing plates are summed. The sum signal of both BPMs is then recorded by a digital oscilloscope that is triggered at \SI{12}{\hertz} and stores 20 frames in total. We use amplifiers with an input impedance of $R=\SI{50}{\ohm}$ close to the BPM devices to establish well-defined signal properties before summing. Neglecting parasitic properties such as resonances, the system of a BPM feeding a resistor can be modeled as a first-order highpass filter driven by a current source~\cite{forck_bpm}. The voltage at the output is:
\begin{equation}
U = Z I_\text{beam} \quad\text{with}\quad Z(\omega) \propto \frac{i \omega RC}{(1 + i \omega RC)} . \label{equation:ohmslaw}
\end{equation}
The pick-up capacitance $C$ is device-dependent, but it is cricital for signal reconstruction only if the maximum frequency of the signal of interest is on the order of $f_\text{cut}=(2\pi RC)^{-1}$ or higher.
With the amplifiers mounted directly on the BPM feedthroughs, the parasitic capacitance can be assumed to be negligible, but the transfer function was not measured in situ.
Our analysis conservatively assumes $f_\text{cut}=\SI{100}{\mega\hertz}$ for the ion BPM and $RC = \infty$ for the electron BPM, the latter being operated without amplifiers.
Since this transfer impedance acts like a differentiator at low frequencies, shortening the bunch length and, correspondingly, increasing the current slope increases the peak amplitude of the signal; the gain structure must be chosen such that the respective peak voltage is handled without clipping under all circumstances.

\begin{figure}[htb]
\includegraphics{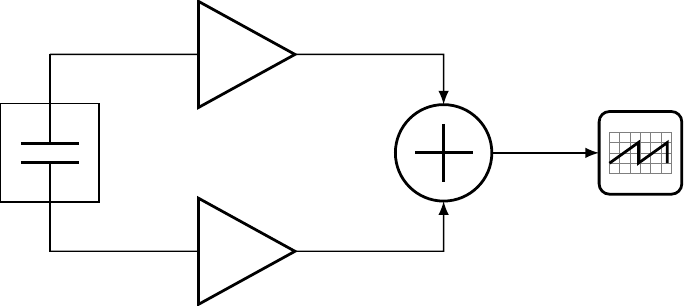}
% \tikzsetnextfilename{fig-bpm-setup}
% \begin{tikzpicture}
% [
% ]
% \ctikzset{capacitors/scale=0.7}
% \draw (-0.5, -0.5) rectangle (0.5, 0.5);
% \draw (0, -1) to [capacitor] (0, 1);
% 
% \draw (4, 0) node[adder] (adder1) {};
% 
% \draw[-latex] (0, -1) to[amp] (4, -1) -- (adder1.south);
% \draw[-latex] (0, 1) to[amp] (4, 1) -- (adder1.north);
% 
% \draw (6, 0) node[oscopeshape] (scope) {};
% %\draw (adder1.east) to[lowpass, l_=\SI{40}{\mega\hertz}] (6.5, 0) -- (scope.west);
% \draw[-latex] (adder1.east) -- (scope.west);
% \end{tikzpicture}
\caption{\label{fig:bpm_setup}Schematic model of the BPM data acquisition setup.}
\end{figure}

We observe the transverse cooling process by measuring the horizontal ion beam profile with an ionization profile monitor (IPM). The device is described in detail in \cite{IPM_Xie2020}.

\subsection{Top-level timing setup}

To prepare an ion beam for pulsed cooling, we first accumulate ions in the ring up to the desired beam current using the standard CSRm accumulation procedure, which takes about \SI{10}{\second} and is accompanied by DC operation of the electron cooler. The resulting beam is unbunched. After accumulation and an additional delay of \SI{2}{\second}, the cooling beam is switched off to let the ion beam heat up for \SI{3}{\second}. The RF system then starts ramping up the cavity voltage, eventually bunching the ion beam. An equilibrium is reached about 5 s after the start of heating. At this point, we start recording the BPM signal to determine the initial bunch profile and switch on the pulsed cooling beam after the first two recorded frames. The whole process after accumulation is recorded by a spectrum analyzer connected to the Schottky pickup to ensure proper timing.

Ideally, the acquisition of the transverse profile from the IPM would coincide with that of the BPM signals. However, for reasons specific to the top-level trigger hardware of the ring, the IPM is started at the same time as the spectrum analyzer.

\begin{figure}[htb]
\includegraphics{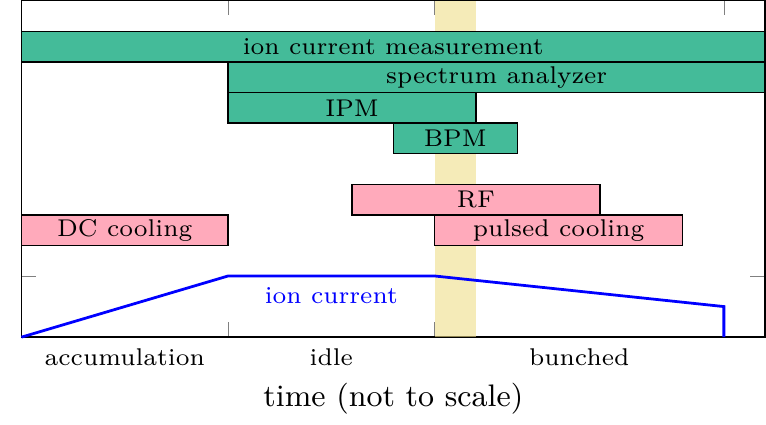}
\caption{\label{fig:toplevel_timing}Sketch of the top-level timing of beam instrumentation components. Components manipulating the beam are shown in red, components measuring its properties in green. The yellow band denotes the time window of interest in which the pulsed cooling process can be observed in all dimensions.}
\end{figure}

The timing scheme for cooling a bunched ion beam is visualized in Fig.~\ref{fig:toplevel_timing}. The setup can also be used to cool a coasting ion beam with a pulsed electron beam, in which case the timing is the same except that the bunching cavity is not powered.

\section{\label{sec:results}Results}

\subsection{\label{sec:schottky_analysis}Spectrum of Schottky pickup signal}

\begin{figure}[htb]
\includegraphics{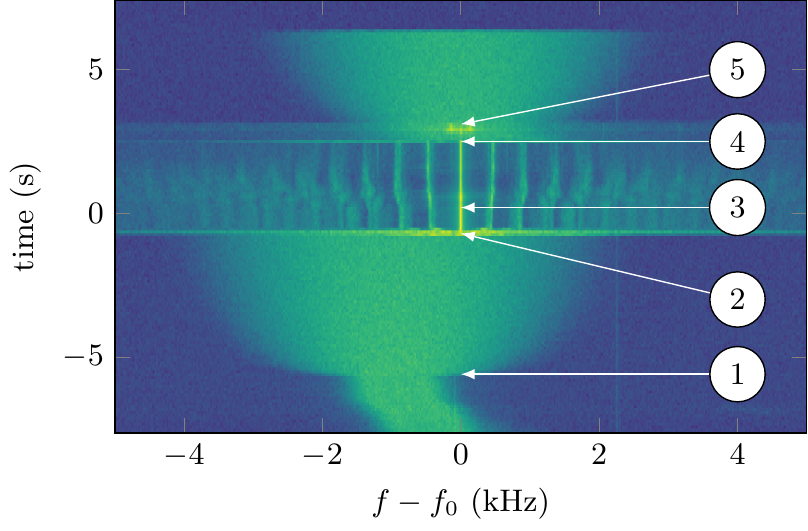}
% \tikzsetnextfilename{fig-spectrogram}
% \begin{tikzpicture}
% [
% 	annotationline/.style={-latex, white},
% 	annotationnode/.style={draw, circle, black, fill=white},
% ]
% \begin{axis}
% [
% 	scale only axis,
% 	width=200pt,
% 	height=125pt,
% 	xlabel={$f - f_0$ (kHz)},
% 	ylabel={time (s)},
% 	enlargelimits=false,
% 	axis on top,
% ]
% \addplot graphics[xmin=-5, xmax=5, ymin=-7.65, ymax=7.4] {data/spectrogram_500.png};
% \draw [annotationline] (axis cs:4, -5.6) node[annotationnode] {1} -- (axis cs:0, -5.6);
% \draw [annotationline] (axis cs:4, -3) node[annotationnode] {2} -- (axis cs:0, -0.7);
% \draw [annotationline] (axis cs:4, 0.2) node[annotationnode] {3} -- (axis cs:0, 0.2);
% \draw [annotationline] (axis cs:4, 2.5) node[annotationnode] {4} -- (axis cs:0, 2.5);
% \draw [annotationline] (axis cs:4, 5) node[annotationnode] {5} -- (axis cs:0, 3.1);
% \end{axis}
% \end{tikzpicture}
\caption{\label{fig:spectrogram}Example of a spectrogram of the 17\textsuperscript{th} revolution harmonic and its synchrotron sidebands measured by the Schottky pickup at $f_0 = \SI{3.256}{\mega\hertz}$. $V_\text{RF,set} = \SI{1}{\kilo\volt}$, electron bunch length \SI{500}{\nano\second}. The color corresponds to a logarithmic scale of spectral power density. $t=0$ denotes the beginning of BPM data acquisition. (1) DC cooler is switched off. (2) Beam starts being bunched. (3) Start of pulsed cooling. (4) RF is switched off. (5) End of pulsed cooling.}
\end{figure}

We use the spectrum from the longitudinal Schottky pickup to determine the synchrotron frequency and, subsequently, the RF voltage.
An example of the evolution of this spectrum as a function of time is given in Fig.~\ref{fig:spectrogram}.
Given that its acquisition time extends from the end of accumulation to beyond the beam dump event, the spectrogram can also be used as a consistency check of the top-level timing.

Synchrotron motion creates sidebands around each harmonic of the revolution frequency with a spacing of the synchrotron frequency $f_\text{S}$~\cite{forck_schottky}.
By measuring this spacing, the RF voltage can be calculated using equation \ref{eq:rf_voltage}~\cite{2014arXiv1404.0927H}:
\begin{equation}
V_\text{RF,calc} = \left(\frac{f_\text{S}}{f_\text{rev}}\right)^2 \frac{2\pi \beta^2 E}{h \eta q}
\label{eq:rf_voltage}
\end{equation}
Here, $f_\text{rev}$ is the revolution frequency, $h=2$ the harmonic number, $\eta=0.952$ the phase slip factor, and $E/q=\SI{3.2}{\giga\volt}$ the total energy of the projectile (including its rest energy) divided by its total charge.
The resulting values of the RF voltage are given in Table~\ref{tab:real_voltage}.
It is evident that the calibration of the set values is inaccurate; in the following sections, the calculated values will be used instead.

\begin{table}[!htbp]%[H] add [H] placement to break table across pages
\caption{RF voltage derived from the measured synchrotron frequency.\label{tab:real_voltage}}
\begin{ruledtabular}
\begin{tabular}{rrr}
$V_\text{RF,set}$ (V) & $f_\text{S}$ (Hz) & $V_\text{RF,calc}$ (V) \\
\hline
600 & $287 \pm 3$ & $252 \pm 5$ \\
800 & $397 \pm 2$ & $483 \pm 6$ \\
1000 & $466 \pm 2$ & $664 \pm 6$ \\
1200 & $537 \pm 1$ & $882 \pm 4$ \\
1500 & $633 \pm 2$ & $1222 \pm 8$ \\
2000 & $753 \pm 2$ & $1730 \pm 9$
\end{tabular}
\end{ruledtabular}
\end{table}

\subsection{\label{sec:bpm_analysis}Analysis of BPM signals}

\begin{figure}[htb]
\includegraphics{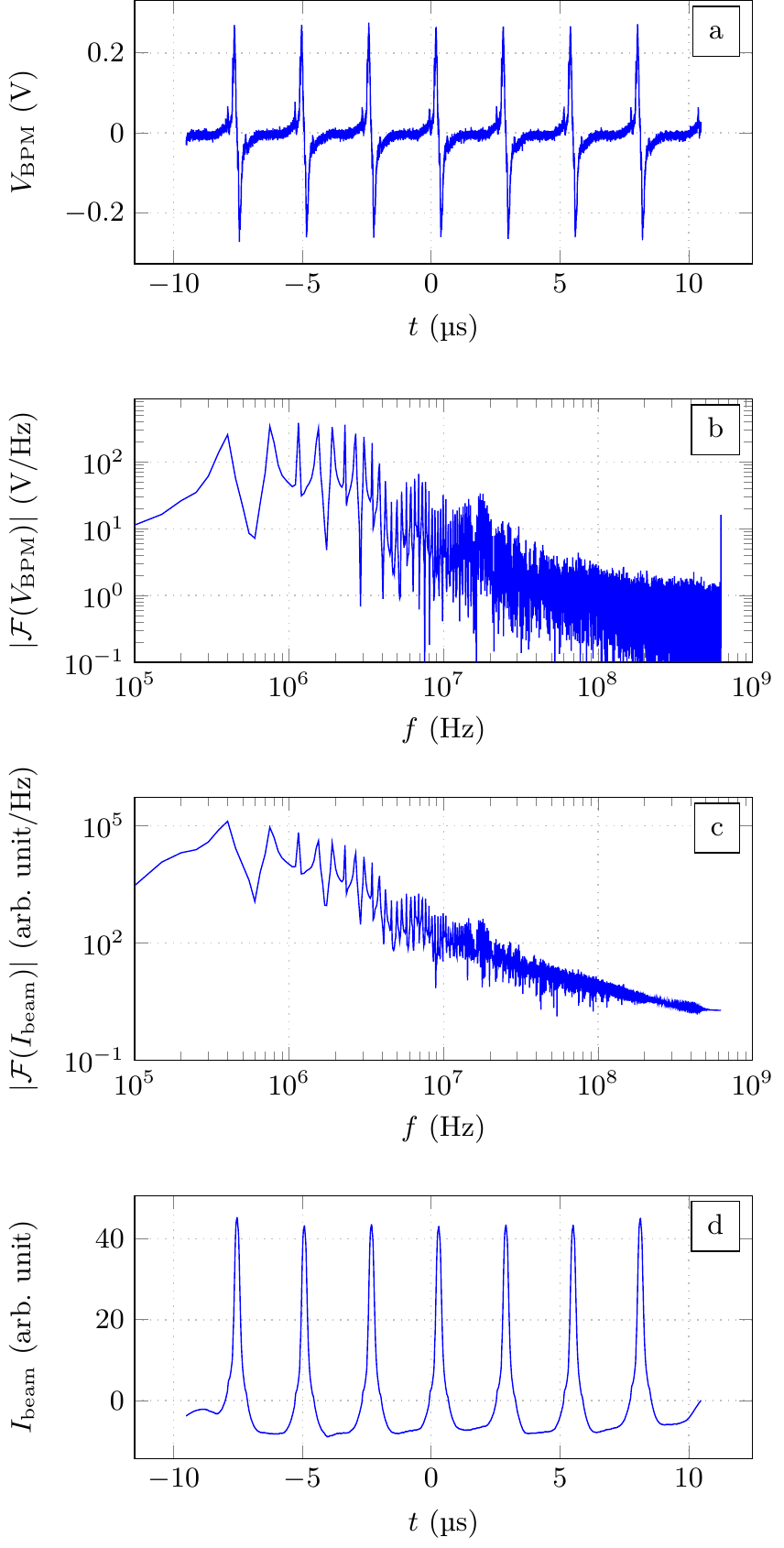}
% \tikzsetnextfilename{fig-bpm-analysis}
% \begin{tikzpicture}
% \begin{groupplot}
% [
% 	group style={group size=1 by 4, vertical sep=15mm},
% 	height=4.5cm,
% 	width=240pt,
% 	xmajorgrids=true,
% 	ymajorgrids=true,
% 	major grid style={dotted},
% 	/pgfplots/every axis y label/.style={at={(0, 0.5)}, xshift=-35pt, rotate=90},
% ]
% \nextgroupplot[xlabel={$t$ (\si{\micro\second})}, ylabel={$V_\text{BPM}$ (V)}]
% \addlegendimage{empty legend};\addlegendentry{\vphantom{fg}a};
% \addplot+[mark=none] table [x expr={\thisrowno{0}*1e6}, y index=1] {data/ibpm_example_unfiltered.dat};
% \nextgroupplot[xmode=log, ymode=log, xmin=1e5, xmax=1e9, ymin=1e-1, xlabel={$f$ (Hz)}, ylabel={$\left|\mathcal{F}(V_\text{BPM})\right|$ (V/Hz)}]
% \addlegendimage{empty legend};\addlegendentry{\vphantom{fg}b};
% \addplot+[mark=none] table [x index=0, y index=1] {data/ibpm_example_unfilteredfft.dat};
% \nextgroupplot[xmode=log, ymode=log, xmin=1e5, xmax=1e9, ymin=1e-1, xlabel={$f$ (Hz)}, ylabel={$\left|\mathcal{F}(I_\text{beam})\right|$ (arb.~unit/Hz)}]
% \addlegendimage{empty legend};\addlegendentry{\vphantom{fg}c};
% \addplot+[mark=none] table [x index=0, y index=1] {data/ibpm_example_filteredfft.dat};
% \nextgroupplot[xlabel={$t$ (\si{\micro\second})}, ylabel={$I_\text{beam}$ (arb.~unit)}]
% \addlegendimage{empty legend};\addlegendentry{\vphantom{fg}d};
% \addplot+[mark=none] table [x expr={\thisrowno{0}*1e6}, y index=1] {data/ibpm_example_filtered.dat};
% \end{groupplot}
% \end{tikzpicture}
\caption{\label{fig:bpm_analysis}Example of the procedure for processing BPM data frames (\SI{500}{\nano\second} electron bunches; $V_\text{RF,set} = \SI{1}{\kilo\volt}$; frame \#8). (a) Raw BPM sum signal as recorded by the oscilloscope during a single trigger event. (b) Frequency-domain representation. (c) Spectrum of filtered signal transformed to beam current. (d) Beam current in the time domain.}
\end{figure}

The output signal of the BPM, being insensitive to the transverse beam properties in this setup, allows for determination of the longitudinal beam profile as follows: First, the digitized BPM signal is windowed with a broad window that mitigates edge effects in the Fourier transform while only affecting the outer 10\,\% of the frame (Tukey; $\alpha=0.1$).
The signal is then transformed to the frequency domain using FFT, converted to beam current using eq.~\ref{equation:ohmslaw}, and transformed back to the time domain.
A 4\textsuperscript{th}-order zero-phase lowpass filter at \SI{200}{\mega\hertz} is used to remove broadband noise and resonances that are unrelated to the beam signal. An example of the transformation process is shown in Fig.~\ref{fig:bpm_analysis}.

The time-domain representation of the beam current shows significant unphysical background at frequencies below the bunch frequency.
As can be seen in the spectrum, this background cannot be filtered out completely in the digital domain because the limited acquisition time per frame causes a poor frequency resolution at the low end. However, the effect of the background on the measured bunch shape can be mitigated by performing a linear correction on each bunch.
Because the oscilloscope is triggered by a slow external signal that is not synchronized with the RF, there is no way to determine which physical bunch corresponds to which recorded bunch. Therefore, we treat the two stored bunches as one with double the revolution frequency so that each frame of data contains four or five bunches that can be averaged for further analysis.
The effect of beam cooling on the bunch shape is negligible during the time span of one frame, and the same is true for synchrotron and betatron motion.
Because the phase of the peaks is (pseudo-)random with respect to the acquisition trigger, the center of each peak is determined by the statistical mean of the signal windowed around the estimate of the respective peak center. Slight fluctuations of this value resulting from asymmetric changes of the bunch shape are unavoidable but inconsequential.

\begin{figure}[htb]
\includegraphics{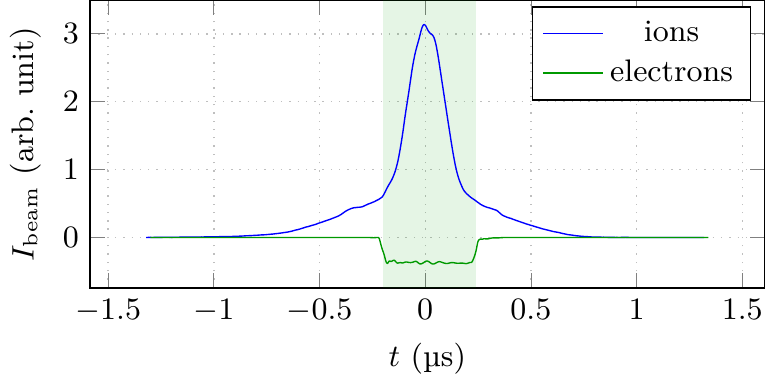}
% \tikzsetnextfilename{fig-single-bunch-analysis}
% \begin{tikzpicture}
% [
% 	electrons/.style={fill=green!60!black, fill opacity=.1, draw=none},
% ]
% \begin{axis}
% [
% 	height=4.5cm,
% 	width=240pt,
% 	xmajorgrids=true,
% 	ymajorgrids=true,
% 	major grid style={dotted},
% 	xlabel={$t$ (\si{\micro\second})},
% 	ylabel={$I_\text{beam}$ (arb.~unit)},
% ]
% \input{data/reg_bpm_edgestikz_500_1_frame8.dat}
% \addplot+[mark=none] table [x expr={\thisrowno{0}*1e6}, y index=1] {data/reg_bpm_ibunch_500_1_frame8.dat};
% \addlegendentry{ions}
% \addplot+[mark=none, green!60!black] table [x expr={\thisrowno{0}*1e6}, y expr={\thisrowno{1}*(-3)}] {data/reg_bpm_ebunch_500_1_frame8.dat};
% \addlegendentry{electrons}
% \end{axis}
% \end{tikzpicture}
\caption{\label{fig:single_bunch_analysis}Example of a frame reduced to a single bunch after corrections and averaging. The electron time axis is shifted to compensate for the time of flight between the BPMs. Parameters as in Fig.~\ref{fig:bpm_analysis}. Note that the visible irregularities in the ion bunch shape are not random and appear in all bunches within the frame.}
\end{figure}

The electron beam current is reconstructed in a similar way, but the length and position of the electron bunches can be determined with greater certainty because their shape stays constant over time. As the rise and fall times are short compared to the total pulse duration, we define the bunch length as the time between the midpoints of the edges and, correspondingly, the bunch center as the point halfway between the edges, neglecting any asymmetries in the shape.
Figure~\ref{fig:single_bunch_analysis} shows an example of the bunch overlap after reconstruction.

\begin{figure}[htb]
\includegraphics{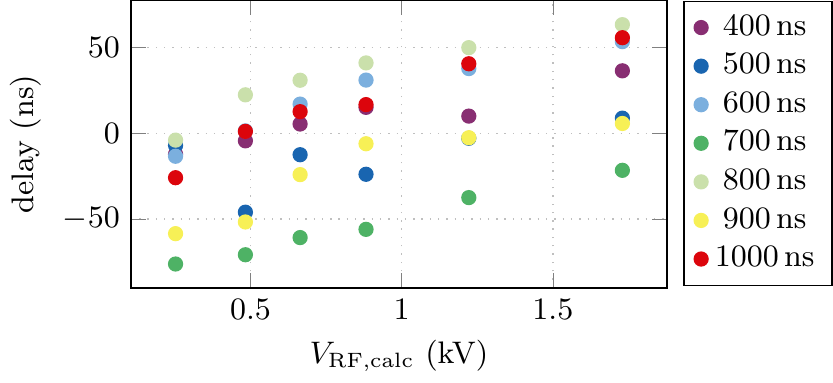}
% \tikzsetnextfilename{fig-bunch-delay-analysis}
% \begin{tikzpicture}
% \begin{axis}
% [
% 	xmajorgrids=true,
% 	ymajorgrids=true,
% 	major grid style={dotted},
% 	cycle list name=TolRainbow7 cycle,
% 	width=200pt,
% 	height=4.5cm,
% 	ylabel={delay (ns)},
% 	xlabel={$V_\text{RF,calc}$ (kV)},
%         legend pos=outer north east,
% ]
% \addplot+[only marks] table[x index=0, y expr={\thisrowno{1}*1e9}] {data/delay_400_frame0.dat}; \addlegendentry{\SI{400}{\nano\second}}
% \addplot+[only marks] table[x index=0, y expr={\thisrowno{1}*1e9}] {data/delay_500_frame0.dat}; \addlegendentry{\SI{500}{\nano\second}}
% \addplot+[only marks] table[x index=0, y expr={\thisrowno{1}*1e9}] {data/delay_600_frame0.dat}; \addlegendentry{\SI{600}{\nano\second}}
% \addplot+[only marks] table[x index=0, y expr={\thisrowno{1}*1e9}] {data/delay_700_frame0.dat}; \addlegendentry{\SI{700}{\nano\second}}
% \addplot+[only marks] table[x index=0, y expr={\thisrowno{1}*1e9}] {data/delay_800_frame0.dat}; \addlegendentry{\SI{800}{\nano\second}}
% \addplot+[only marks] table[x index=0, y expr={\thisrowno{1}*1e9}] {data/delay_900_frame0.dat}; \addlegendentry{\SI{900}{\nano\second}}
% \addplot+[only marks] table[x index=0, y expr={\thisrowno{1}*1e9}] {data/delay_1000_frame0.dat}; \addlegendentry{\SI{1000}{\nano\second}}
% \end{axis}
% \end{tikzpicture}
\caption{\label{fig:bunch_delay_analysis}Delay between ion bunch center and electron bunch center as a function of RF voltage and electron bunch length; evaluated at the start of cooling.}
\end{figure}

This analysis allows a consistency check of the actual bunch overlap and length as a function of parameter settings. The delay between the bunches is shown in Fig.~\ref{fig:bunch_delay_analysis}.
The delay shows random deviations as a function of the bunch length setting, which is a result of manual adjustment between experimental runs. The systematic change in bunch delay as a function of RF voltage is assumed to be a property of the RF signal from which the pulse synchronization trigger is derived.

\begin{figure}
\includegraphics{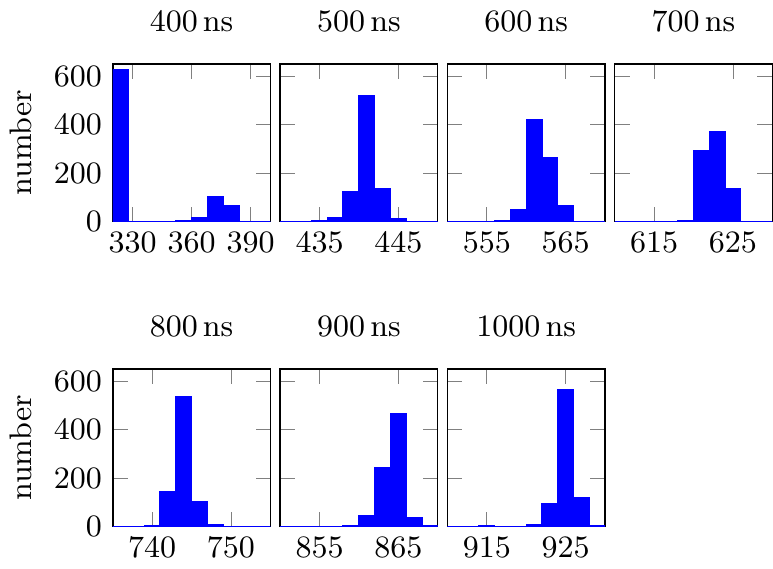}
\caption{\label{fig:bunch_length_histogram}Histogram of observed electron bunch lengths (in ns), one plot per setting. Due to a deficiency of the circuit triggering the grid pulser, the central values of the distributions differ from the set point by varying amounts. In the \SI{400}{\nano\second} case, two different bunch lengths are created.}
\end{figure}

Figure \ref{fig:bunch_length_histogram} shows the distribution of measured electron bunch lengths; a statistical analysis is listed in table~\ref{tab:bunch_length_table}.
The bunch length is generally stable, having an RMS jitter level comparable to the temporal resolution of the BPM measurement.
However, the central value of the distribution differs from the nominal value by a varying amount, which is attributed to a deficiency of the signal transmission circuit driving the grid pulser.
The \SI{400}{\nano\second} case is peculiar in that this same issue results in two different bunch lengths being generated at random.

\begin{table}[htbp]%[H] add [H] placement to break table across pages
\caption{Analysis of the observed electron bunch length distributions (see Fig.~\ref{fig:bunch_length_histogram})\label{tab:bunch_length_table}}
\begin{ruledtabular}
\begin{tabular}{ccc}
nominal (ns) & measured mean (ns) & measured $\sqrt{\sigma^2}$ (ns) \\
\hline
400 & 333 & 20.7 \\
500 & 441 & 1.4 \\
600 & 562 & 1.4 \\
700 & 622 & 1.6 \\
800 & 744 & 1.3 \\
900 & 864 & 1.3 \\
1000 & 925 & 2.2
\end{tabular}
\end{ruledtabular}
\end{table}

\subsection{\label{sec:ipm_analysis}Analysis of IPM signals}

\begin{figure}[htb]
\includegraphics{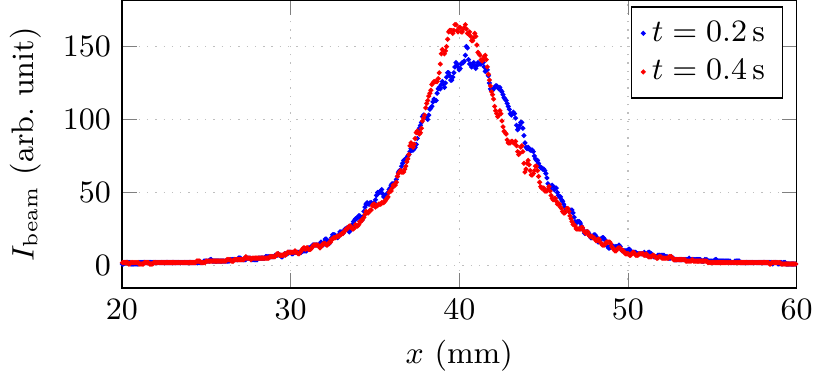}
% \tikzsetnextfilename{fig-ipm-example}
% \begin{tikzpicture}
% [
% 	ipmdata/.style={only marks, mark=*, mark size=.5pt}
% ]
% \begin{axis}
% [
% 	height=4.5cm,
% 	width=240pt,
% 	xmajorgrids=true,
% 	ymajorgrids=true,
% 	major grid style={dotted},
% 	xlabel={$x$ (\si{\milli\meter})},
% 	ylabel={$I_\text{beam}$ (arb.~unit)},
% 	xmin=20, xmax=60,
% ]
% \addplot+[ipmdata] table [x index=0, y index=1] {data/ipm_1kv_500ns_frame-2.dat}; \addlegendentry{$t=\SI{0.2}{\second}$}
% \addplot+[ipmdata] table [x index=0, y index=1] {data/ipm_1kv_500ns_frame-1.dat}; \addlegendentry{$t=\SI{0.4}{\second}$}
% \end{axis}
% \end{tikzpicture}
\caption{\label{fig:ipm_example}Example of the evolution of the horizontal profile during pulsed cooling. Parameters as in Fig.~\ref{fig:bpm_analysis}.}
\end{figure}

The transverse beam profile projected to the horizontal axis is recorded as a function of time with an exposure time of $\SI{200}{\milli\second}$ per frame.
Because of the way the trigger initiating the acquisition is set up and the total number of frames is fixed, most of the data are taken during beam preparation prior to pulsed cooling. Only two frames are available that show the cooling process. An example of the data is shown in Fig.~\ref{fig:ipm_example}.

Because one transverse axis is averaged away during acquisition, the recorded profile has both high resolution and low noise, so its statistical properties can be analyzed without any preprocessing.
However, the observed total intensity systematically depends on the beam shape for unknown reasons, so the results must be interpreted with care.
The data are not corrupted by digital clipping or similar effects, but nonlinearities on the analog side cannot be ruled out.

\subsection{\label{sec:coastingcooling}Cooling properties with a coasting ion beam}

\begin{figure}[htb]
\includegraphics{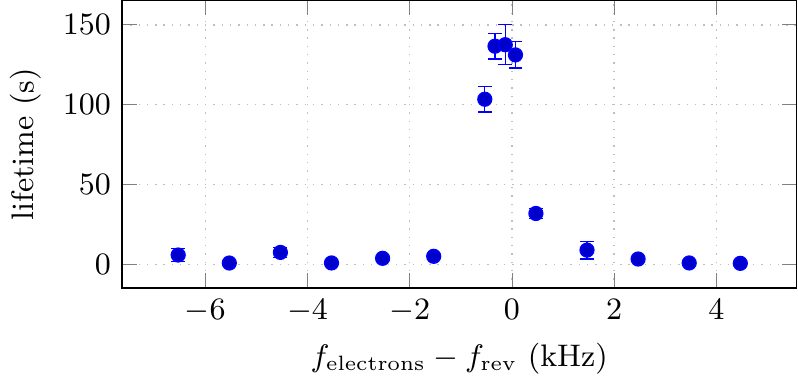}
% \tikzsetnextfilename{fig-coasting-lifetime}
% \begin{tikzpicture}
% \begin{axis}
% [
% 	width=240pt,
% 	height=4.5cm,
% 	xmajorgrids=true,
% 	ymajorgrids=true,
% 	major grid style={dotted},
% 	xlabel={$f_\text{electrons} - f_\text{rev}$ (\si{\kilo\hertz})},
% 	ylabel={lifetime (\si{\second})},
% ]
% \addplot+[only marks, error bars/y dir=both, error bars/y explicit] table [x expr={\thisrowno{0}-191.53}, y index=1, y error index=2] {data/coasting_lifetime.dat};
% \end{axis}
% \end{tikzpicture}
\caption{\label{fig:coasting_lifetime}Dependence of the ion beam lifetime on the frequency mismatch between ion and electron pulses. The error bars result from averaging over multiple runs.}
\end{figure}

As a first check of the cooling dynamics, we investigated the process of a coasting ion beam being cooled by the pulsed electron beam with the RF cavity in the ring switched off. The frequency of the electron pulse repetition signal was varied around the ion revolution frequency. As shown in Fig.~\ref{fig:coasting_lifetime}, an anomalous reduction of the lifetime of the stored ion beam was observed at mismatched frequencies, while in the case of equal frequencies, the lifetime is equal to that observed with a DC electron beam.
As will be shown in section \ref{sec:spacecharge}, this particle loss is caused by space charge kicks perturbing the phase space in both the transverse and the longitudinal plane.

\begin{figure}[htb]
\includegraphics{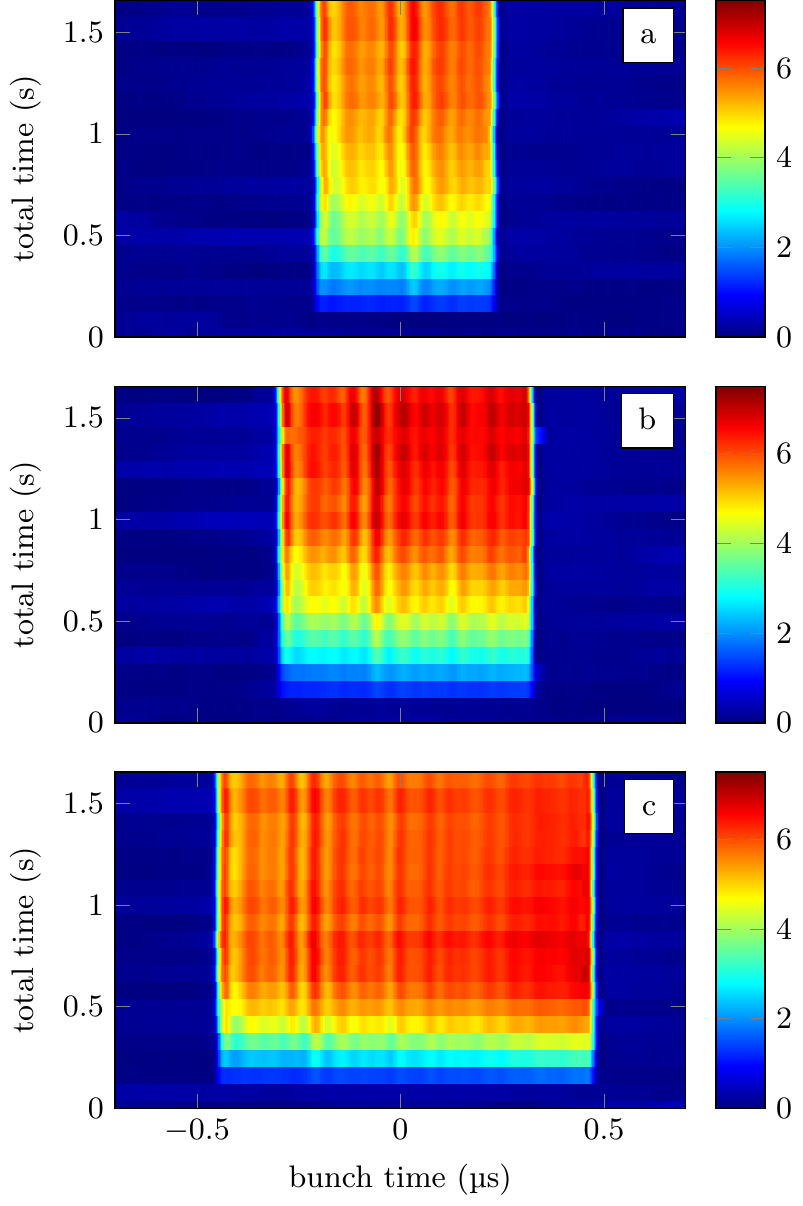}
\caption{\label{fig:coasting_bunching}Evolution of the longitudinal beam profile of the initially unbunched ion beam during the cooling process measured at different electron pulse lengths. (a) \SI{500}{\nano\second}, (b) \SI{700}{\nano\second}, (c) \SI{900}{\nano\second}. The color scale corresponds to an arbitrary unit of beam current and is normalized to an independent measurement of the respective DC current prior to cooling for the sake of comparability. The apparent increase in total intensity over time results from the inability of the BPM to detect the DC component of the beam. The non-uniform intensity distribution of the cooled ion bunch is identical among independent experiments and was also observed in our previous simulations, particularly for shorter electron pulses~\cite{ZHAO2018219}.}
\end{figure}

In the case of frequency-matched pulses, another interesting phenomenon is the bunching of the coasting beam as a result of the bucket created by the space charge field of the electron beam (called \enquote{grouping effect} in prior work \cite{ZHAO2018219, Mao:COOL2017-TUP15}). The evolution of the longitudinal ion beam profile in this case is shown in Fig.~\ref{fig:coasting_bunching}. It is evident that the ions are captured in a longitudinal space corresponding to the electron bunch length. Because the cooling force eventually reduces their momentum deviation below the bucket height, they cannot escape from the bucket.

\subsection{\label{sec:cooling}Cooling properties with a bunched ion beam}

In the main part of the experiment, the RF system was switched on, capturing the ions in corresponding buckets after accumulation.
This way, we were able to measure a bunched ion beam being cooled by a pulsed electron beam including synchrotron dynamics.
Using both the longitudinal and transverse beam profiles recorded during pulsed cooling with different RF voltages and electron bunch lengths, we can derive the respective cooling rates as a function of these parameters.

\begin{figure}[htb]
\includegraphics{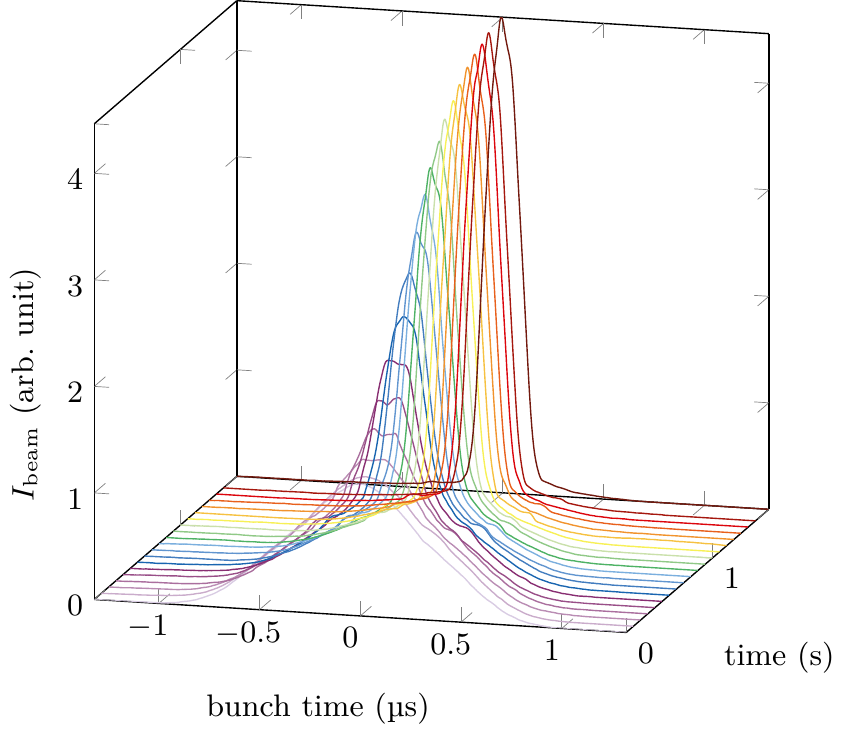}
\caption{\label{fig:bunch_evolution}Example of the evolution of the longitudinal bunch profile. Parameters as in Fig.~\ref{fig:bpm_analysis}. The cooling beam is switched on at $t = \SI{0.1}{\second}$.}
\end{figure}

An example of the temporal evolution of the longitudinal bunch shape during cooling is shown in Fig.~\ref{fig:bunch_evolution}.
We observe an overall reduction of the bunch length as a function of time. Regardless of the initial shape, the resulting profile is non-Gaussian with irregularities both near the core and in the tails. While the overall shape of the profile is quantifiable by computing the sample moments, the details of these features vary so much across the data set that a dedicated systematic measurement will be warranted if they are to be fully understood.

For lack of an analytic model of the bunch shape, we quantify its properties by extracting the following quantities directly from the intensity $I(t)$, with $t$ being an equidistantly spaced, discrete time axis:
\begin{align}
Q &= \sum I & \quad(\propto \text{bunch charge}) \\
\hat{t} &= \frac{\sum I(t) t}{\sum I(t)} & (\text{bunch center}) \\
\sigma^2 &= \frac{\sum I(t) (t-\hat{t})^2}{\sum I(t)} & (\text{variance}) \\
\kappa &= \frac{\sum I(t) (t-\hat{t})^4}{\sigma^4 \sum I(t)} & (\text{kurtosis})
\end{align}
$\sigma = \sqrt{\sigma^2}$ is a measure of the overall width of the distribution. The value of $\kappa$ is 3 for a Gaussian shape; a higher value indicates a shift of the probability density toward the tails of the distribution.

\begin{figure}[htb]
\includegraphics{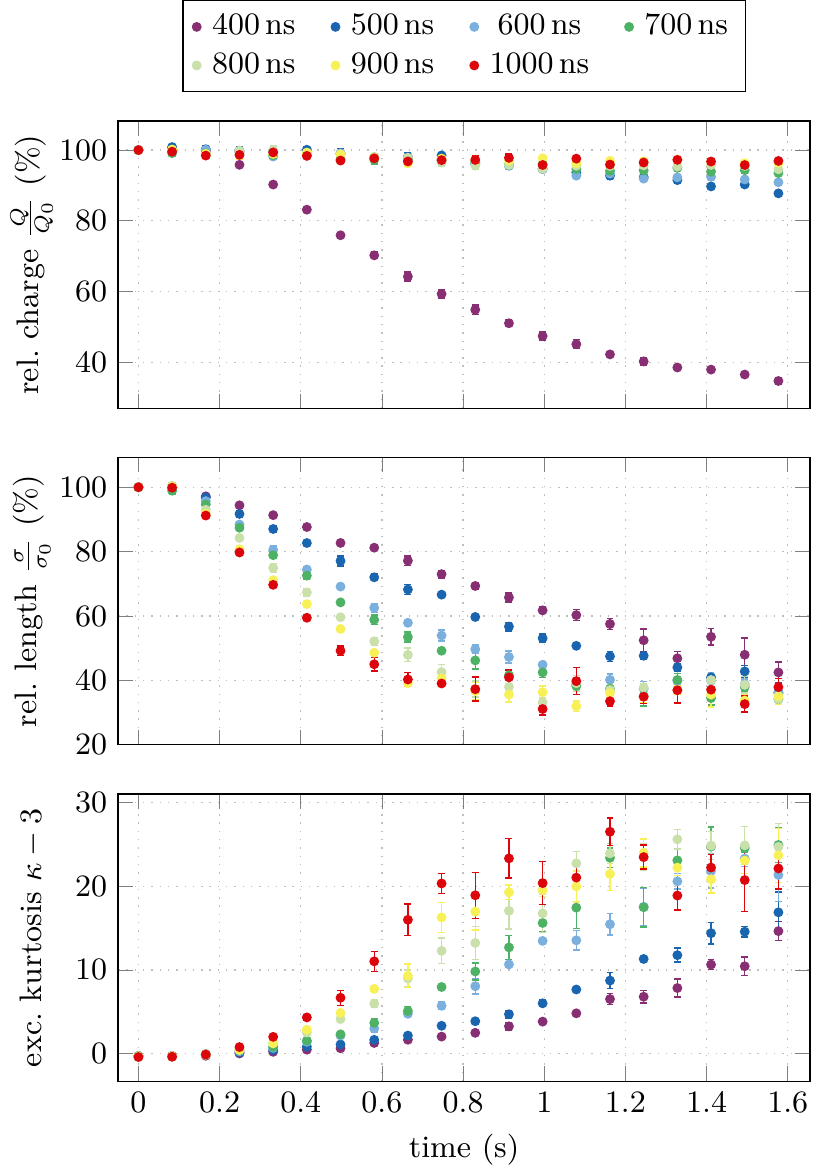}
\caption{\label{fig:bpm_moments_1kv}Evolution of the statistical properties of the longitudinal ion bunch profile as a function of time, averaged over five identical experimental runs per setting. Each color corresponds to one setting of the electron bunch length. The cooling beam is switched on at $t = \SI{0.1}{\second}$. $V_\text{RF,set} = \SI{1}{\kilo\volt}$. The error bars represent the statistical error of the mean.}
\end{figure}

Figure \ref{fig:bpm_moments_1kv} shows the evolution of these quantities at a nominal RF voltage of \SI{1}{\kilo\volt}. The most salient aspect of this result is the particle loss taking place at an electron bunch length of \SI{400}{\nano\second}. This effect is visible at any RF voltage and can be attributed to single particles randomly being subject to both longitudinal and transverse heating as a result of the varying space-charge-induced focusing force brought about by the electron bunch length jitter, which is a unique property of the \SI{400}{\nano\second} setting (see Fig.~\ref{fig:bunch_length_histogram}).
A simulation showing this mechanism is described in section \ref{sec:spacecharge}.

Apart from particle loss, we observe a monotonic dependency between bunch length and cooling rate.
Since the longitudinal current density in the electron bunches is constant, i.e.~changing the bunch length changes only the longitudinal overlap but not the peak current, this result is to be expected as long as the length of the electron bunches does not significantly exceed that of the ion bunches.
Conversely, the evolution of the kurtosis hints at a tendency for longer electron bunches to be detrimental to preserving a Gaussian shape.
While our experiment was carried out at a constant phase between electron and ion bunches (albeit not perfectly, see Fig.~\ref{fig:bunch_delay_analysis}), this result warrants a dedicated measurement of how the evolution of the bunch tails is affected by the phase.

\begin{figure}[htb]
\includegraphics{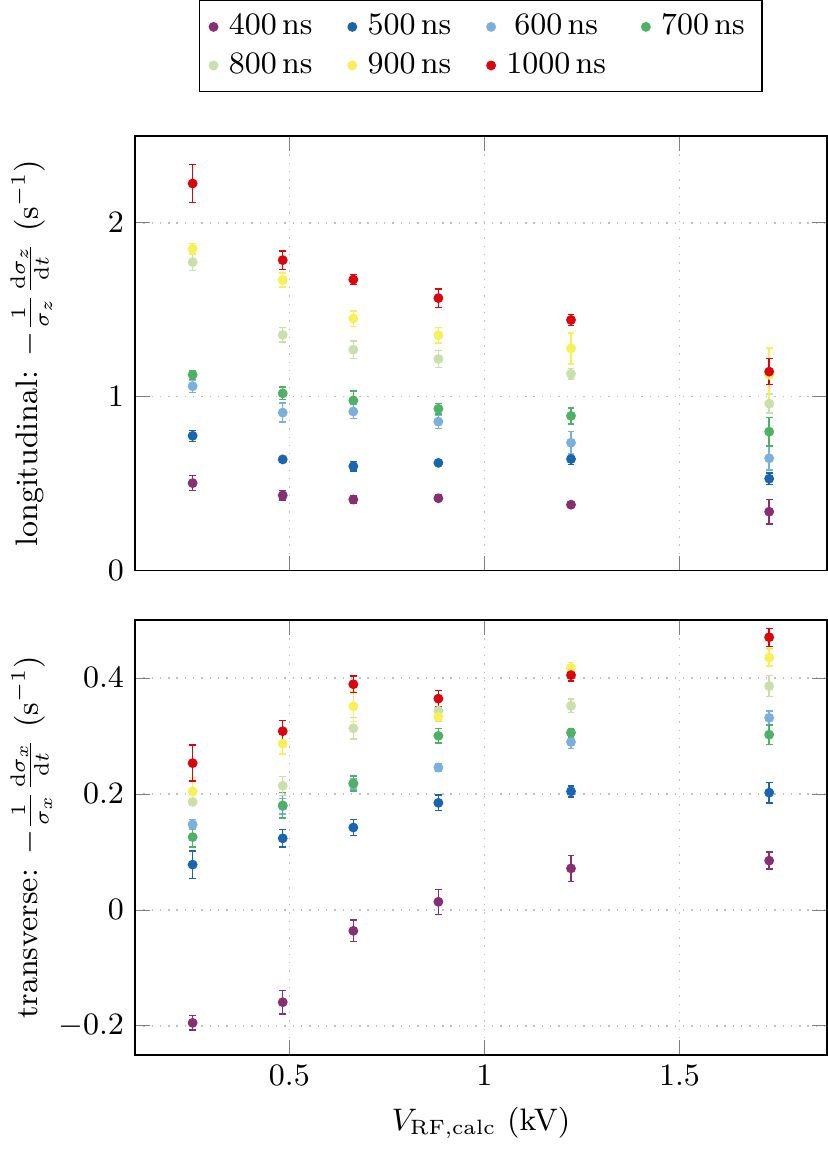}
\caption{\label{fig:bpm_moments_diff}Longitudinal and transverse cooling rates as a function of electron bunch length and RF voltage, calculated in the interval $\SI{0.2}{\second} \leq t \leq \SI{0.4}{\second}$ and averaged over five identical experimental runs per setting. The error bars represent the statistical error of the mean.}
\end{figure}

As shown in section \ref{sec:jspec}, the asymptotic behavior toward the end of the process is predominantly caused by intra-beam scattering (IBS) counteracting the cooling force.
Because this effect does not contribute significantly until the bunches become short, the longitudinal and transverse cooling rates are determined from the evolution of the bunch length or width, respectively, in a region of the curves where no asymptotic behavior is visible.
The relative slope of the $\sigma$ curve is summarized in Fig.~\ref{fig:bpm_moments_diff} for all parameter sets of our experiment.

The longitudinal cooling rate increases monotonically as a function of the electron bunch length, which is to be expected because longer bunches lead to a higher overlap between the bunches on average. It can, however, be observed that increasing the RF voltage leads to the opposite effect even though that, too, increases the overlap. We suspect this reduction in cooling rate with increasing RF voltage to be a result of the broadening of the longitudinal momentum distribution, which causes the average cooling force to decrease despite the higher bunch overlap.
This effect must be especially strong when the overlap is centered around the bunch center, where the momentum deviation of high-amplitude particles is highest by virtue of synchrotron motion.

The transverse cooling rate increases monotonically as a function of both the RF focusing strength and the electron bunch length because it only depends on the longitudinal overlap between the bunches. The transverse overlap is always the same, and stronger RF focusing does not broaden the transverse momentum distribution.

The change in focusing strength due to the bunch-length-dependent space charge potential of the electron beam should also have an effect on the bunch shape but cannot be determined in isolation in this experiment.

\section{\label{sec:jspec}Simulation of cooling and IBS rates}

\begin{table}[!htbp]%[H] add [H] placement to break table across pages
\caption{Cooling simulation parameters\label{tab:jspec_parameters}}
\begin{ruledtabular}
\begin{tabular}{lc}
Transverse electron beam radius & \SI{15}{\milli\meter} \\
Effective cooler length & \SI{3.4}{\meter} \\
Magnetic field & \SI{0.1}{\tesla} \\
Average $\beta_x$ / $\beta_y$ in the cooler & \SI{10}{\meter} / \SI{17}{\meter} \\
Peak electron current (uniform bunch shape) & \SI{30}{\milli\ampere} \\
Transverse electron temperature & \SI{200}{\milli\electronvolt} \\
Longitudinal electron temperature & \SI{6}{\milli\electronvolt} \\
RMS normalized transverse emittance & \SI{0.6}{\milli\meter\milli\radian} \\
RMS long.~ion bunch size $\sigma_z$ & \SI{10.5}{\meter} \\
RMS long.~ion momentum deviation $\sigma_{p_z / p}$ & \num{7e-4}
\end{tabular}
\end{ruledtabular}
\end{table}

To gain confidence in the interpretation of the results, a set of simulations mimicking the experimental conditions was carried out using the code JSPEC~\cite{Zhang:IPAC2016-WEPMW014, jspec}.
In these simulations, the evolution of the bunch shape was computed on a turn-by-turn basis by applying theoretical models for both the cooling force and the heating effect caused by intra-beam scattering (IBS).
RF focusing is parametrized by the synchrotron frequency.

Table \ref{tab:jspec_parameters} shows the input parameters used. The longitudinal profile of the simulated electron bunches is rectangular; since the true bunch length in the experiment is different from the nominal setting as shown in Fig.~\ref{fig:bunch_length_histogram}, the averages of the respective measured lengths are used for the simulation.
While the code can consider an arbitrary temporal alignment between the bunches, a systematic study of the effect of the alignment was not part of the experiment, and the small offsets shown in Fig.~\ref{fig:bunch_delay_analysis} are considered negligible for the purpose of this study.
Because the cooling rate calculation is particularly sensitive to beam parameters that are hard to determine accurately, e.g.~the electron beam temperature, the resulting absolute numbers only serve as an order-of-magnitude estimate; however, the qualitative behavior and the relative dependence on the cooling bunch length can still be meaningfully compared to the experimental data.

\begin{figure}[htb]
\includegraphics{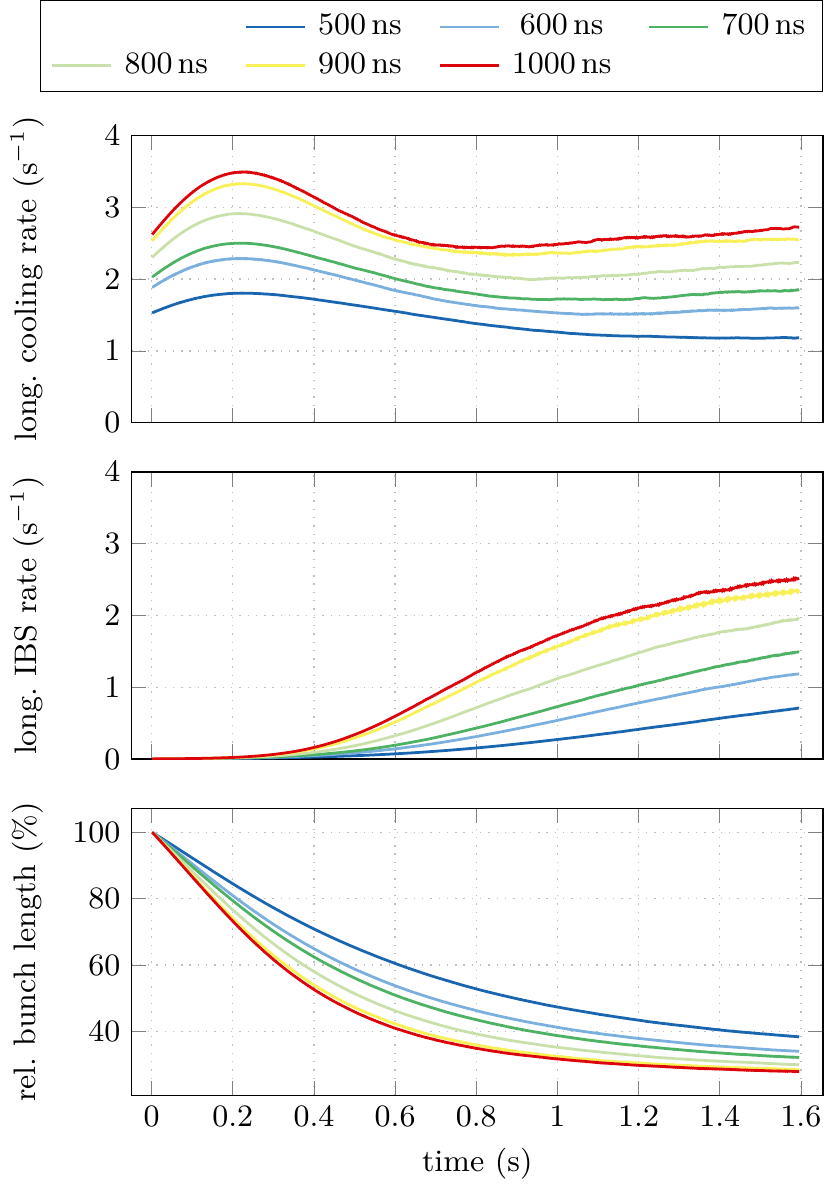}
\caption{\label{fig:jspec_rates}Simulation of cooling rate, IBS rate, and ion bunch length with parameter sets similar to the experiment. Note that the definition of the rates used here refers to the emittance rather than the bunch length. The \SI{400}{\nano\second} case was omitted because this type of simulation cannot reproduce the effect of bunch length jitter. $f_\text{S}=\SI{466}{\hertz}$.}
\end{figure}

Figure \ref{fig:jspec_rates} shows the evolution of the cooling rate, the IBS rate, and the resulting bunch length from the simulation at $f_\text{S} = \SI{466}{\hertz}$, corresponding to $V_\text{RF,set}=\SI{1}{\kilo\volt}$ as in Fig.~\ref{fig:bpm_moments_1kv}.
Here, we concentrate on the longitudinal behavior because the simulation is assumed to model the transverse cooling force inaccurately.
The bunch length evolution observed in the experiment is qualitatively reproduced.
While the beginning of the process is dominated by cooling, the contribution of IBS increases significantly as a result of the three-dimensional compression of the ion bunch, eventually canceling out the cooling effect and resulting in an equilibrium.

Table \ref{tab:jspec_comparison} shows a comparison of the measured and simulated bunch length reduction rates, the latter being determined from the slope of the bunch length curve in the same way as in Fig.~\ref{fig:bpm_moments_diff}.
While the simulation tends to overestimate the cooling rate toward lower electron bunch lengths, the overall behavior is in reasonable agreement considering the sensitivity to the models and beam parameters.

\begin{table}[!htbp]%[H] add [H] placement to break table across pages
\caption{Comparison between simulated and measured bunch length reduction rates. $T_\text{set}$ and $T_\text{exp}$ are the nominal and measured electron bunch lengths, respectively. $R_\text{exp}$ is the longitudinal bunch length reduction rate as shown in Fig.~\ref{fig:bpm_moments_diff}; $R_\text{sim}$ is the corresponding simulation result.\label{tab:jspec_comparison}}
\begin{ruledtabular}
\begin{tabular}{ccccc}
$T_\text{set}$ (ns)& $T_\text{exp}$ (ns) & $R_\text{exp}$ (\si{\per\second})& $R_\text{sim}$ (\si{\per\second}) & $R_\text{sim} / R_\text{exp}$ \\
\hline
500 & 441 & 0.60 & 0.89 & 1.48 \\
600 & 562 & 0.92 & 1.17 & 1.28 \\
700 & 623 & 0.98 & 1.23 & 1.25 \\
800 & 744 & 1.27 & 1.45 & 1.14 \\
900 & 864 & 1.45 & 1.69 & 1.16 \\
1000 & 925 & 1.67 & 1.77 & 1.06
\end{tabular}
\end{ruledtabular}
\end{table}

\section{\label{sec:spacecharge}Simulation of space-charge effects}

The particle loss taking place in the runs with \SI{400}{\nano\second} electron bunch length was initially attributed to a transverse heating mechanism due to uneven space charge kicks caused by bunch length jitter. This mechanism had already been described both theoretically and with simulations for the case of LEReC, where, assuming a significant phase jitter, the heating is not negligible but still small enough not to cause considerable particle losses \cite{gangwang, Blaskiewicz:NAPAC2016-WEA3IO01}.
While the same basic principle applies to our experiment, the particles are nonrelativistic and space charge forces comparatively high as a result.
The possibility of the longitudinal dynamics also being affected should therefore not be ruled out.

In an effort to assess the magnitude of the effect, we carried out a tracking simulation including the space charge kicks from the electron beam in the cooler.
The simulation code tracks the 6-dimensional phase space coordinates of randomly generated single ions through the linear transport matrices of the ring and applies the longitudinal cavity kick at each revolution. The cooling section is discretized by a finite number of drifts, in which the electron charge distribution is placed according to the relative phase of the ion when it reaches each individual piece of drift. Changes in velocity and, thus, relative time of flight are accounted for. The charge distribution is then discretized on a cartesian, three-dimensional grid and the Coulomb force computed.
Collective effects of the ion beam are assumed to be negligible.
This model includes neither cooling nor IBS so that the space charge effect can be observed in isolation.
The assumptions made as input to the simulation are listed in table~\ref{tab:simulation_params}.
While the simplicity of the model causes some of them to be arbitrary, they do not affect the qualitative outcome significantly.

\begin{table}[!htbp]%[H] add [H] placement to break table across pages
\caption{Parameters for tracking simulations and theoretical estimates\label{tab:simulation_params}}
\begin{ruledtabular}
\begin{tabular}{lc}
Peak electron current $I_\text{e}$ & \SI{30}{\milli\ampere} \\
Bunch length $T_\text{bunch}$ & \SI{320}{\nano\second} or \SI{370}{\nano\second} \\
Electron current rise/fall time & \SI{30}{\nano\second} \\
Transverse electron beam radius $a_\text{e}$ & \SI{15}{\milli\meter} \\
Cooler length $L_\text{cool}$ & \SI{3.4}{\meter} \\
Ion ring betatron tunes $Q_{x,y}$ & 3.62 / 2.61 \\
Ion ring RF voltage (see table \ref{tab:real_voltage}) & \SI{664}{\volt} \\
Transverse aperture limit $(\sqrt{x^2+y^2})_\text{max}$ & \SI{50}{\milli\meter} \\
Number of ions & 400 \\
RMS transverse ion bunch radius $\sigma_{x,y}$ & \SI{5}{\milli\meter} \\
RMS transverse ion momentum $\sigma_{p_{x,y}} / p_z$ & \num{1e-3} \\
RMS long.~ion bunch size $\sigma_z$ & \SI{7.5}{\meter} \\
RMS long.~ion momentum deviation $\sigma_{p_z / p}$ & \num{5e-4}
\end{tabular}
\end{ruledtabular}
\end{table}

\begin{figure}[!htb]
\includegraphics{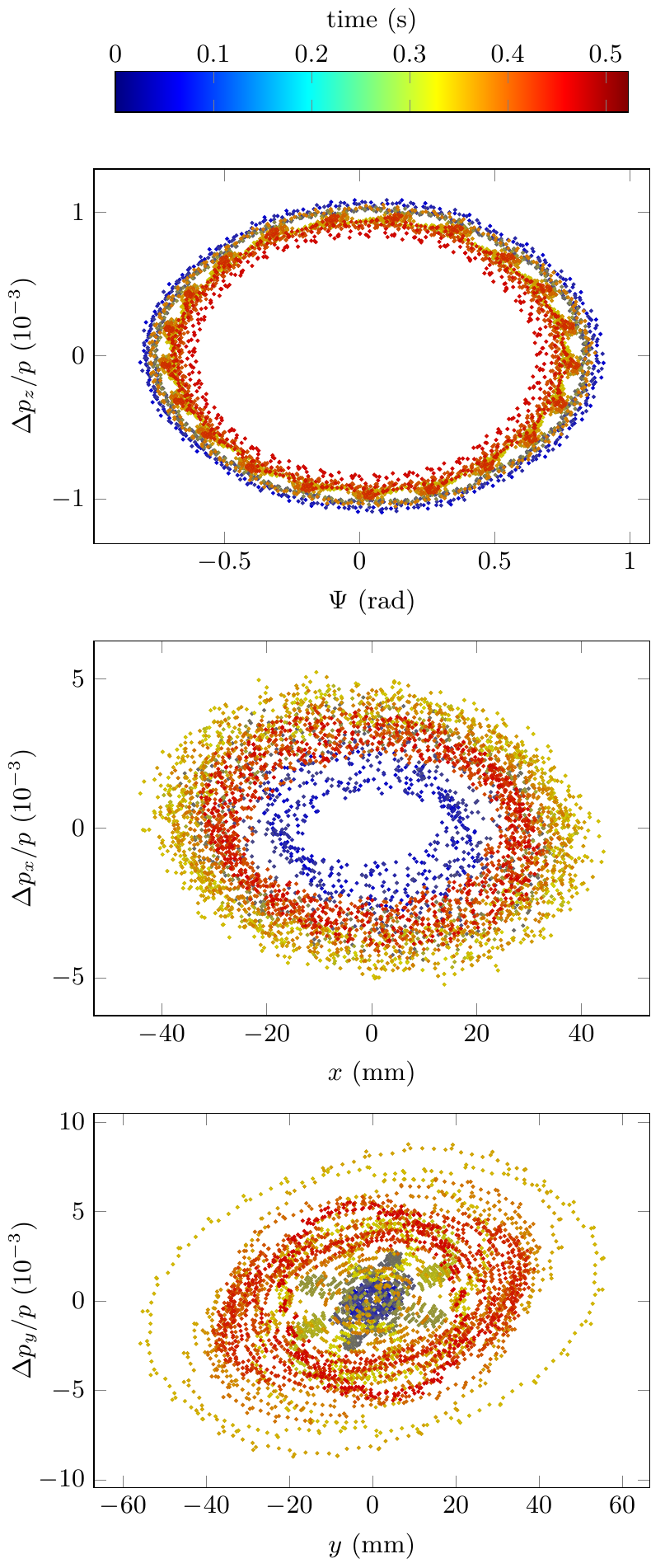}
\caption{\label{fig:jitter_particle}Phase space trajectory of one example particle placed randomly in 6-d phase space and subjected to cooling bunches with bunch length jitter. The perturbation of the longitudinal motion is inconsequential, while fluctuations of the transverse action eventually lead to a collision with a limiting aperture. The plot only shows every 20\textsuperscript{th} revolution for clarity, whereas collisions with the aperture are checked after each machine element.}
\end{figure}

\begin{figure}[t]
\includegraphics{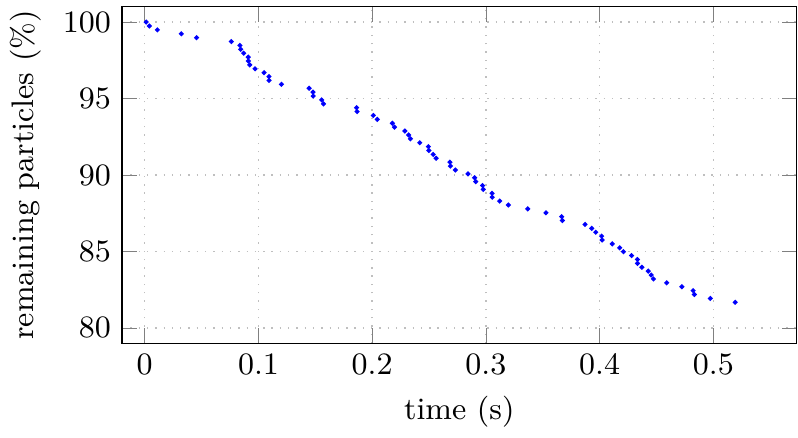}
% \tikzsetnextfilename{fig-jitter-lossrate}
% \begin{tikzpicture}
% \begin{axis}
% [
% 	width=240pt,
% 	height=5cm,
% 	xlabel={time (s)},
% 	ylabel={remaining particles (\%)},
% 	xticklabel style={/pgf/number format/.cd,fixed,precision=2},
% 	xmajorgrids=true,
% 	ymajorgrids=true,
% 	major grid style={dotted},
% 	ymin=79,
% 	ymax=101,
% 	xmin=-0.02,
% ]
% \addplot+[only marks, mark size=.5pt] table [x expr={\thisrowno{0}/191500.0}, y index=1] {data/jitter_lossrate.dat};
% \end{axis}
% \end{tikzpicture}
\caption{\label{fig:jitter_lossrate}Relative particle loss with bunch length jitter as a function of time. The initial sample contains 400 particles randomly distributed according to table~\ref{tab:simulation_params}.}
\end{figure}

To simulate the bunch length jitter present in the \SI{400}{\nano\second} experiment, we assume the starting edge of any bunch to randomly arrive early by \SI{50}{\nano\second} with a probability of 1/3 to mimic the distribution in Fig.~\ref{fig:bunch_length_histogram}.
The phase space trajectory of an example particle subjected to these conditions is shown in Fig.~\ref{fig:jitter_particle}.
While the action in the longitudinal plane is visibly affected, there is no indication of a spontaneous blow-up or instability that would explain the loss of the particle by itself.
The transverse plane, however, shows a significant deformation of the trajectory over time.
Under the simplifying assumption that any particle with a transverse displacement of $\sqrt{x^2+y^2} > \SI{50}{\milli\meter}$ is lost, which is checked after each optical element, we obtain the number of lost particles as a function of time; the result is shown in Fig.~\ref{fig:jitter_lossrate}.
The observed loss of 18\,\% in \SI{0.5}{\second} is in reasonable agreement with the experiment considering the simplifications.
The same simulation without bunch length jitter gives no particle loss.

These results allow us to make a comparison with existing theoretical models: It is shown in \cite{gangwang} that a randomly distributed electron bunch arrival time of variance $\left<\delta_t^2\right>$ results in an exponential ion emittance growth, assuming longitudinally Gaussian-shaped electron bunches of length $\sigma_\text{e}$.
While the case at hand is slightly different in that our bunches are rectangular and their length varies, it is similar enough to make a comparison as a rough consistency check.
The growth time in either the $x$ or $y$ plane is given by \cite{gangwang}:
\begin{align}\label{tau_theo}
\tau_{\text{theo},x/y} = \frac{\exp(1)}{8 \pi^2 f_\text{rev}} \frac{\sigma_\text{e}^2}{\left<\delta_t^2\right>} \frac{1}{\Delta\nu_{\text{peak},x/y}^2} .
\end{align}
We choose the equivalent Gaussian bunch length $\sigma_\text{e}$ such that the FWHM is equal to the true rectangular bunch length; with the ion radius $r_0$, the number of electrons per bunch $N_\text{e}$, and the arrival time variance $\left<\delta_t^2\right>$ as given by
\begin{align}
r_0 &= \SI{4.5e-19}{\meter}\;\text{for}\;{}^{86}\mathrm{Kr}^{25+} ,\\
N_\text{e} &= \frac{I_\text{e}}{e_0} (T_\text{bunch} + T_\text{rise/fall}) = \num{7.1e10} , \\
\sigma_\text{e} &= \frac{T_\text{bunch}}{2\sqrt{2 \ln 2}} = \SI{161}{\nano\second} , \\
\left<\delta_t^2\right> &= (\SI{20.7}{\nano\second})^2 ,
\end{align}
we obtain \cite{gangwang}:
\begin{align}
\Delta \nu_{\text{peak},x/y} &= \frac{N_\text{e} r_0 L_\text{cool}}{(2\pi)^2 Q_{x/y} f_\text{rev} \sqrt{2\pi} \sigma_\text{e} \gamma^3 \beta^2 a_\text{e}^2} \\
&= \begin{pmatrix}\num{4.1e-3}\\\num{5.6e-3}\end{pmatrix} , \\
\tau_{\text{theo},x/y} &= \begin{pmatrix}\SI{0.66}{\second}\\\SI{0.34}{\second}\end{pmatrix} .
\end{align}

\begin{figure}[htb]
\includegraphics{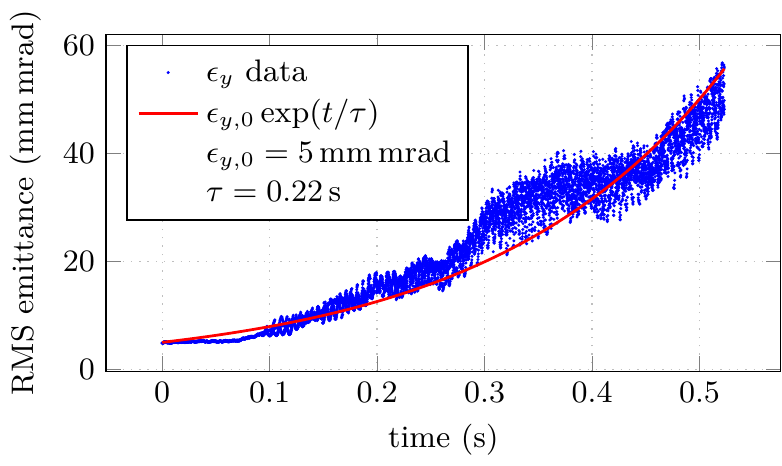}
% \tikzsetnextfilename{fig-jitter-emittance-y}
% \begin{tikzpicture}
% \begin{axis}
% [
% 	width=240pt,
% 	height=5cm,
% 	xlabel={time (s)},
% 	ylabel={RMS emittance (mm\,mrad)},
% %	xticklabel style={/pgf/number format/.cd,fixed,precision=2},
% 	xmajorgrids=true,
% 	ymajorgrids=true,
% 	major grid style={dotted},
% %	ymin=0,
% %	ymax=101,
% %	xmin=-0.02,
% 	legend pos=north west,
% 	legend cell align={left},
% 	clip mode=individual,
% ]
% \addplot+[only marks, mark size=.2pt] table [x index=0, y index=1] {data/jitter_emittance_y.dat}; \addlegendentry{$\epsilon_y$ data}
% \addplot+[mark=none, thick] table [x index=0, y index=2] {data/jitter_emittance_y.dat}; \addlegendentry{$\epsilon_{y,0} \exp(t/\tau)$}
% \addlegendimage{empty legend}\addlegendentry{$\epsilon_{y,0} = \SI{5}{\milli\meter\milli\radian}$}
% \addlegendimage{empty legend}\addlegendentry{$\tau = \SI{0.22}{\second}$}
% \end{axis}
% \end{tikzpicture}
\caption{\label{fig:jitter_emittance_y}Growth of the RMS emittance as a result of bunch length jitter as obtained from the tracking simulation. The growth time $\tau$ is a free parameter, while the initial emittance $\epsilon_{y,0}$ is fixed.}
\end{figure}

For comparison with this model, we can extract an emittance growth time from the full simulated ensemble, keeping in mind that the emittance ceases to be well-defined in the face of particle loss and that the heated ions spend considerable time outside of the cooling beam, which is in contrast to the assumptions made in \cite{gangwang}.
The RMS emittance in the $y$ plane (chosen here because of the non-dispersive optics) is calculated from the ensemble phase space via
\begin{equation}\label{eq:emittance}
\epsilon_y = \sqrt{\left<y^2\right> \left<y'^2\right> - \left<y y'\right>^2} ,
\end{equation}
where $y$ denotes the displacement and $y'$ the angle of each particle with respect to the reference orbit.
In this calculation, to avoid discontinuities in the data, lost particles are kept, but their emittance does not grow further.
Figure~\ref{fig:jitter_emittance_y} shows the evolution of the emittance.
While the data do not show a strictly exponential time dependence, applying an exponential fit allows us to extract a growth time of $\tau = \SI{0.22}{\second}$ for the sake of comparison with the model-based value $\tau_\text{theo}$.
The latter is higher by a factor of \num{1.5}.

Considering the number of assumptions going into both the simulation and the growth rate estimate, the discrepancy between the results is not surprising.
However, the qualitative agreement between the methods demonstrates their basic applicability and consistency.
While the results we obtained explain the particle loss observed in our experiment and imply that the bunch phase and length ought to be held as constant as possible in bunched-beam devices, the consequences of the effect are expected to be less severe for relativistic beams as space charge forces diminish at $\gamma \gg 1$.
The theory presented in \cite{gangwang} appears to be able to make a good prediction of the heating time in that case.

\section{\label{sec:conclusion}Conclusion}

We have demonstrated cooling of both coasting and bunched ion beams using a pulsed electron beam from a conventional cooler, allowing studies of bunched-beam cooling with relatively little technical effort and investment.
Even though the parameters of our experiment are far away from those of the most likely application of bunched-beam cooling, i.e.~high-energy protons \cite{eic}, the basic physics are similar.

The experimental result has been shown to be in reasonable albeit not perfect agreement with simulations.
Improving the input parameters of these simulations would require considerable beam diagnostics efforts.
While the level of agreement we achieved allows us to be confident in the overall scheme, it does not exclude the possibility of there being small heating effects in the set-up that have not been fully explored.

We have shown the observed particle loss to likely result from random transverse heating caused by uneven space charge kicks. This observation hints at the practical importance of maintaining the cooling bunch length and phase with high accuracy. An important implication is that sweeping the cooling bunch longitudinally on a short time scale is likely not an option for future machines. However, considering that the cooling time of conceivable high-energy coolers is between many minutes and an hour \cite{eic}, applying such a sweeping technique on a long time scale may still be feasible.

% Specify following sections are appendices. Use \appendix* if there
% only one appendix.
%\appendix
%\section{}

% If you have acknowledgments, this puts in the proper section head.
\begin{acknowledgments}
The authors would like to thank all staff of the CSR operation group at IMP.
We would also like to thank John Musson, Edith Nissen, Christiana Wilson, and Jianxun Yan of Jefferson Lab for supporting the experimental runs, and Robert McKeown and Mike Spata of Jefferson Lab for encouragement and support.

This material is based upon work supported by the U.S. Department of Energy, Office of Science, Office of Nuclear Physics under contract DE-AC05-06OR23177.

This experiment is supported by the International Partnership Program of Chinese Academy of Sciences, Grant No.~113462KYSB20170051, and the National Natural Science Foundation of China, No.~11575264.
\end{acknowledgments}

% Create the reference section using BibTeX:
\bibliography{impcool_prab}

\end{document}